\documentclass{emulateapj}
\usepackage[]{natbib, graphicx}

\begin{document}

\title{The Role of Mergers in Early-type Galaxy Evolution and Black Hole Growth}

\shorttitle{Mergers \& AGN in Early-type Galaxies}
\shortauthors{Schawinski et al.}
\slugcomment{Accepted for publication in the Astrophysical Journal Letters}

\author{
Kevin Schawinski,\altaffilmark{1,2,3}
Nathan Dowlin,\altaffilmark{2,3}
Daniel Thomas,\altaffilmark{4}
C. Megan Urry\altaffilmark{2,3,5} and 
Edward Edmondson\altaffilmark{4}
}

\altaffiltext{1}{Einstein Fellow}
\altaffiltext{2}{Department of Physics, Yale University, New Haven, CT 06511, U.S.A.}
\altaffiltext{3}{Yale Center for Astronomy and Astrophysics, Yale University, P.O. Box 208121, New Haven, CT 06520, U.S.A.}
\altaffiltext{4}{Institute of Cosmology and Gravitation, University of Portsmouth, Mercantile House, Hampshire Terrace, Portsmouth, PO1 2EG, UK}
\altaffiltext{5}{Department of Astronomy, Yale University, New Haven, CT 06511, U.S.A.}

\email{kevin.schawinski@yale.edu}

\def\Chandra{\textit{Chandra}}
\def\XMM{\textit{XMM-Newton}}
\def\Swift{\textit{Swift}}

\def\OI{[\mbox{O\,{\sc i}}]~$\lambda 6300$}
\def\OIII{[\mbox{O\,{\sc iii}}]~$\lambda 5007$}
\def\SII{[\mbox{S\,{\sc ii}}]~$\lambda \lambda 6717,6731$}
\def\NII{[\mbox{N\,{\sc ii}}]~$\lambda 6584$}

\def\Ha{{H$\alpha$}}
\def\Hb{{H$\beta$}}

\def\NIIHa{[\mbox{N\,{\sc ii}}]/H$\alpha$}
\def\SIIHa{[\mbox{S\,{\sc ii}}]/H$\alpha$}
\def\OIHa{[\mbox{O\,{\sc i}}]/H$\alpha$}
\def\OIIIHb{[\mbox{O\,{\sc iii}}]/H$\beta$}

\def\Ebmv{E($B-V$)}
\def\LOIII{$L[\mbox{O\,{\sc iii}}]$}
\def\Ledd{${L/L_{\rm Edd}}$}
\def\LOIIIs4{$L[\mbox{O\,{\sc iii}}]$/$\sigma^4$}
\def\LOIIIMbh{$L[\mbox{O\,{\sc iii}}]$/$M_{\rm BH}$}
\def\Mbh{$M_{\rm BH}$}
\def\Msigma{$M_{\rm BH} - \sigma$}
\def\Ms{$M_{\rm *}$}
\def\Msun{$M_{\odot}$}
\def\Msunyr{$M_{\odot}yr^{-1}$}

\def\ergs{$~\rm ergs^{-1}$}
\def\kms{$~\rm kms^{-1}$}

\begin{abstract}
Models of galaxy formation invoke the major merger of gas-rich progenitor galaxies as the trigger for significant phases of black hole growth and the associated feedback that suppresses star formation to create red spheroidal remnants. However, the observational evidence for the connection between mergers and active galactic nucleus (AGN) phases is not clear. We analyze a sample of low-mass early-type galaxies known to be in the process of migrating from the blue cloud to the red sequence via an AGN phase in the green valley. Using deeper imaging from SDSS Stripe 82, we show that the fraction of objects with major morphological disturbances is high during the early starburst phase, but declines rapidly to the background level seen in quiescent early-type galaxies by the time of substantial AGN radiation several hundred Myr after the starburst. This observation empirically links the AGN activity in low-redshift early-type galaxies to a significant merger event in the recent past. The large time delay between the merger-driven starburst and the peak of AGN activity allows for the merger features to decay to the background and hence may explain the weak link between merger features and AGN activity in the literature. 
\end{abstract}

\keywords{galaxies: interactions; galaxies: evolution; galaxies: formation galaxies: Seyfert; galaxies: active}

\section{Introduction}
\label{sec:intro}

In 1988, \citeauthor{1988ApJ...325...74S} proposed a direct link between major mergers and quasar activity: specifically, that major mergers between gas-rich galaxies fuel a substantial starburst; eventually some of the gas reaches the black hole, triggering an accretion event; and the resulting energy output from the quasar, such as radiation and outflows, sweeps up the remaining gas, thus suppressing further star formation and creating a passively evolving spheroidal remnant. 

The hierarchical nature of the $\Lambda$CDM cosmology suggests that mergers of haloes and galaxies are one of the main drivers of evolution. Inspired by the \cite{1988ApJ...325...74S} scenario, simulations of individual galaxy mergers show that major mergers can radically transform the progenitors by destroying any inbound disks and fueling major starbursts and heating part of the gas content \citep[\textit{e.g.},][]{1992ApJ...393..484B,1996ApJ...471..115B, 2006MNRAS.373.1013C}. More recent works have added the second component by invoking some version of quasar feedback to suppress star formation and establishing various observed scaling relations  \citep[\textit{e.g.},][]{2005Natur.433..604D,2005ApJ...620L..79S, 2005MNRAS.361..776S, 2006ApJS..163....1H, 2008ApJS..175..390H, 2008ApJS..175..356H, 2008MNRAS.391..481S, 2009ApJ...690..802J}. This scenario, in which mergers trigger starbursts and AGN phases, now underpins the current generation of theoretical models as a framework for the co-evolution of galaxies and black holes.

However, the observational evidence connecting these various phases is not clear. Many studies find that incidence of merger features in AGN host galaxies does not seem to be significantly enhanced over the general population \citep[\textit{e.g.},][]{1998ApJ...496...93D, 1998ApJS..117...25M, 2001AJ....122.2243S,  2007ApJ...660L..19P, 2009MNRAS.397..623G, 2009ApJ...691..705G, 2010MNRAS.401.1552D}, while a minority find some evidence for such an enhancement \citep[\textit{e.g.},][]{1986ApJ...311..526H,1996AJ....111..696K,2008ApJ...679.1047K, 2008ApJ...674...80U}. At the same time, recent observational work has questioned the degree to which mergers are involved in both the triggering of starbursts and the subsequent quenching of star formation and formation of red spheroidal galaxies, as the rate of major mergers accounts for all the mass growth in red galaxies since $z \sim 1$ \citep{2007ApJ...665L...5B, 2009ApJ...697.1369B}.

\begin{figure*}
\begin{center}

\includegraphics[angle=270, width=\textwidth]{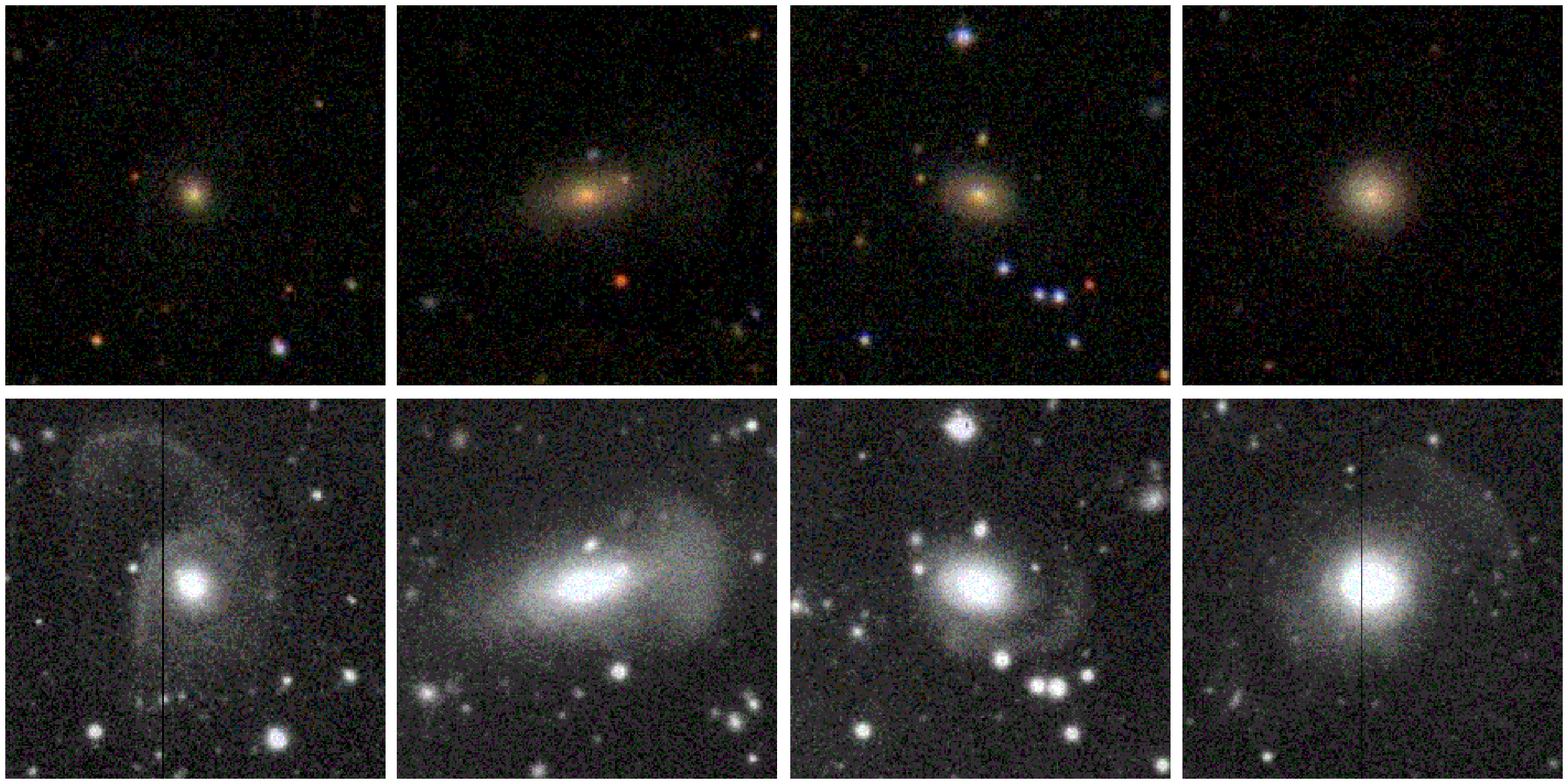}
\caption{Comparison of regular SDSS images to Stripe 82 images of four early-type galaxies that clearly show features indicating a recent merger in the deeper data. In the \textit{top} row, we show the regular SDSS $gri$-composite images, while in the \textit{bottom} row, we show the corresponding deeper Stripe 82 co-added $r$-band images. Features that are readily apparent in the Stripe 82 images are not visible in the regular SDSS images.\label{fig:sdss_vs_stripe82_merger}}

\includegraphics[angle=270, width=\textwidth]{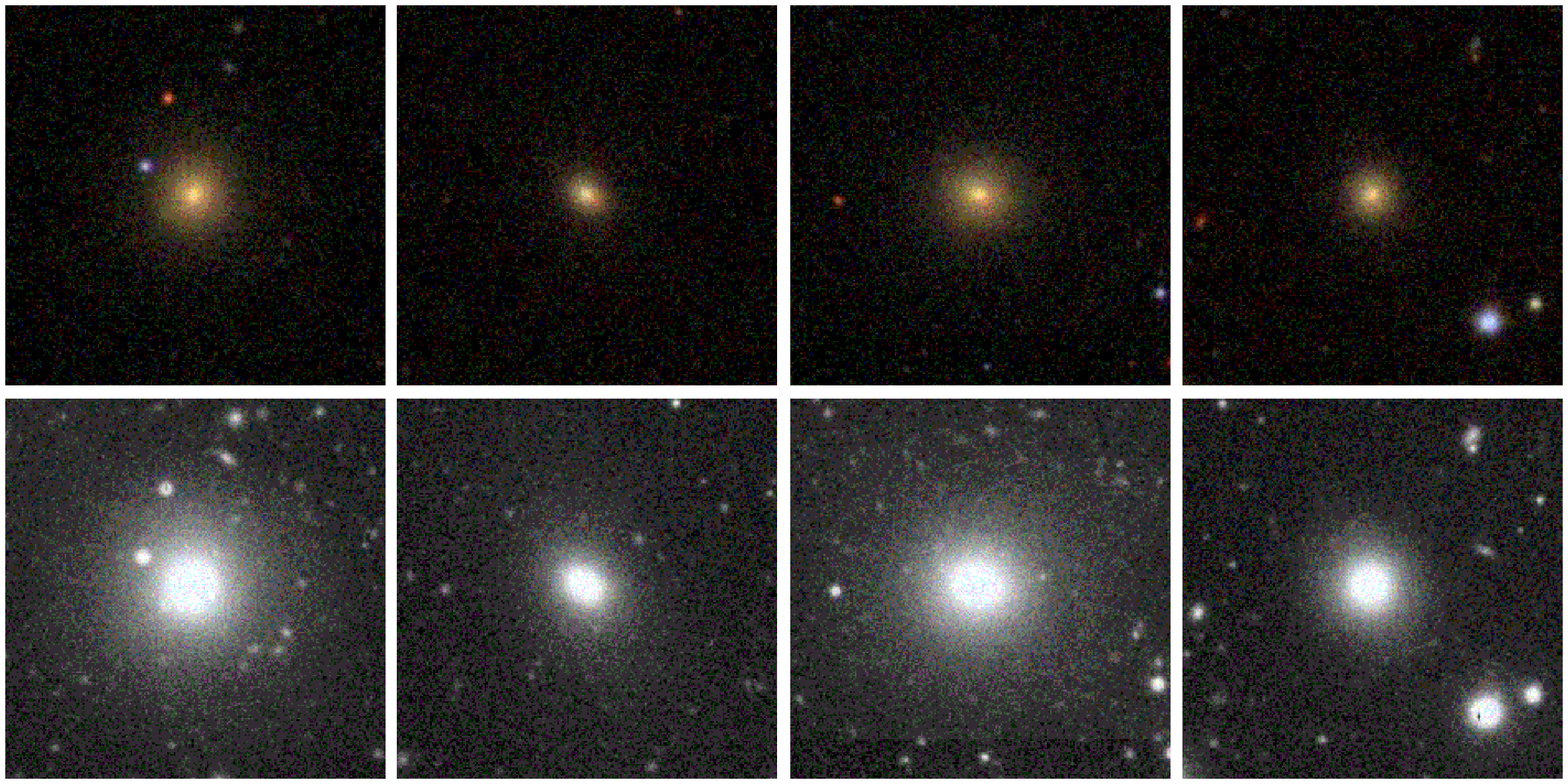}
\caption{Same as Figure \ref{fig:sdss_vs_stripe82_merger}, but for early-type galaxies that do not show any merger features in the corresponding deeper Stripe 82 images. \label{fig:sdss_vs_stripe82_nomerger}}

\end{center}
\end{figure*}

We decided to test this evolutionary scenario using a sample of spheroidal galaxies that are known to be in the process of migrating from the blue cloud to the red sequence via an AGN phase. We can therefore identify the same population in various evolutionary phases during the galaxy transformation process. The migration of early-type galaxies has been traced for low-mass early-type galaxies by \cite{2007MNRAS.382.1415S} using the MOSES sample\footnote{Morphologically Selected Early-types from Sloan, see \cite{2007MNRAS.382.1415S}}. The objects along this sequence are an ideal laboratory to test whether merging activity is in any way related to both the shutdown of star formation -- leading to the migration from blue to red -- and the fueling of the central black hole, which may also be involved in creating a red, passively evolving remnant. 

In this Letter, we test the role of mergers by obtaining significantly deeper imaging data for early-types along the evolutionary sequence of \cite{2007MNRAS.382.1415S}. The original SDSS images used for the morphological classification by eye performed by \cite{2007MNRAS.382.1415S} do not reach sufficiently deep surface brightness limits to detect features indicating a recent merger, such as shells, tidal tails and other large-scale disruptions. If deeper imaging data of the same objects reveals such features, a link is established between merging activity on the one hand, and the shutdown of star formation and black hole fueling on the other. We obtain such deeper images from SDSS Stripe 82.

\section{Analysis}

\subsection{Sample Selection \& Experimental Setup}

At stellar masses around $10^{10}$\Msun, morphological early-type galaxies drawn from the Sloan Digital Sky Survey ($z\sim0.05$), with ongoing activity traced by strong optical emission lines, form a clear evolutionary sequence: (1) early-type galaxies, with current star formation reside in the blue cloud on the color-mass diagram; (2) early-types whose emission line ratios indicate that the output of ionizing photons from star formation and nuclear activity are roughly comparable are at the same mass, but exhibit slightly redder optical colors; (3) objects dominated by nebular emission from a Seyfert AGN cluster in the green valley;  (4) objects with weak LINER\footnote{Low ionization nuclear emission region; \cite{2010MNRAS.402.2187S} argue that the power source of LINER emission in early-type galaxies are evolved stellar populations and not AGN.} emission and (5) quiescent early-types at the low-mass end of the red sequence. The increasingly red optical colors of galaxies along this sequence suggest an evolutionary sequence. A detailed analysis of the stellar populations of objects along this sequence, taking into account both the broad-band UV-optical-near-IR spectral energy distribution, as well as the stellar absorption (Lick) indices confirms that this is a genuine evolutionary sequence \citep{2007MNRAS.382.1415S}.

Since even old red and dead galaxies show signs of mergers in very deep images \citep{2005AJ....130.2647V}, it is important to establish the background level of disturbed galaxies in addition to comparing the incidence of mergers in galaxies as a function of time since the merger. If mergers trigger the migration from blue to red, then as galaxies become redder, the merger signs should become less frequent and less obvious and eventually reach the background level seen in passive red sequence galaxies. We therefore use the quiescent early-types from the MOSES sample to determine this background level of the merger fraction.

\subsection{SDSS and Stripe 82 Imaging Data}

We use both the regular SDSS imaging data from Data Release 4 \citep{2000AJ....120.1579Y, 2006ApJS..162...38A} as well as the deeper Stripe 82 data\footnote{See \texttt{http://www.sdss.org/legacy/stripe82.html}}. SDSS repeatedly imaged an equatorial stripe -- Stripe 82 -- to search for optical transients \citep[\textit{e.g.},][]{2007AJ....134.2236S}. Co-adding the images from multiple epochs has produced a large area of $\sim270 ~\rm{deg}^2$ of imaging approximately 2 magnitudes deeper than an individual SDSS scan. We use these deeper Stripe 82 images to search for evidence for recent merger activity in MOSES early-type galaxies which are not apparent in the regular SDSS images. We provide example images of MOSES early-types from both regular SDSS and Stripe 82 in Figures \ref{fig:sdss_vs_stripe82_merger} and \ref{fig:sdss_vs_stripe82_nomerger}.

\subsection{Merger Fraction Measurement}

We measure the \textit{merger fraction} for each MOSES early-type galaxy by visually inspecting the Stripe 82 image. Each image is randomly displayed and classified by one of us (ND) into objects that do show features indicating a recent merger (such as tidal tails, fans, and other disruptions; see Figure \ref{fig:sdss_vs_stripe82_merger}), and those without such features (see Figure \ref{fig:sdss_vs_stripe82_nomerger}). Each repeat classification of an object was done blind, with no information about any previous classifications. We process the entire sample five times to ensure the consistency of our classifications over time. The merger fraction of a population is calculated by taking the fraction of of objects with merger signatures for each of the five individual series of classification and averaging them. The confidence intervals on the merger fraction are computed using the Binomial distribution appropriate for small number statistics following \cite{1986ApJ...303..336G} except in bins where numbers are large ($N > 300$), where we assume the error to be the Poisson error.

\begin{figure}
\begin{center}

\includegraphics[angle=90, width=0.49\textwidth]{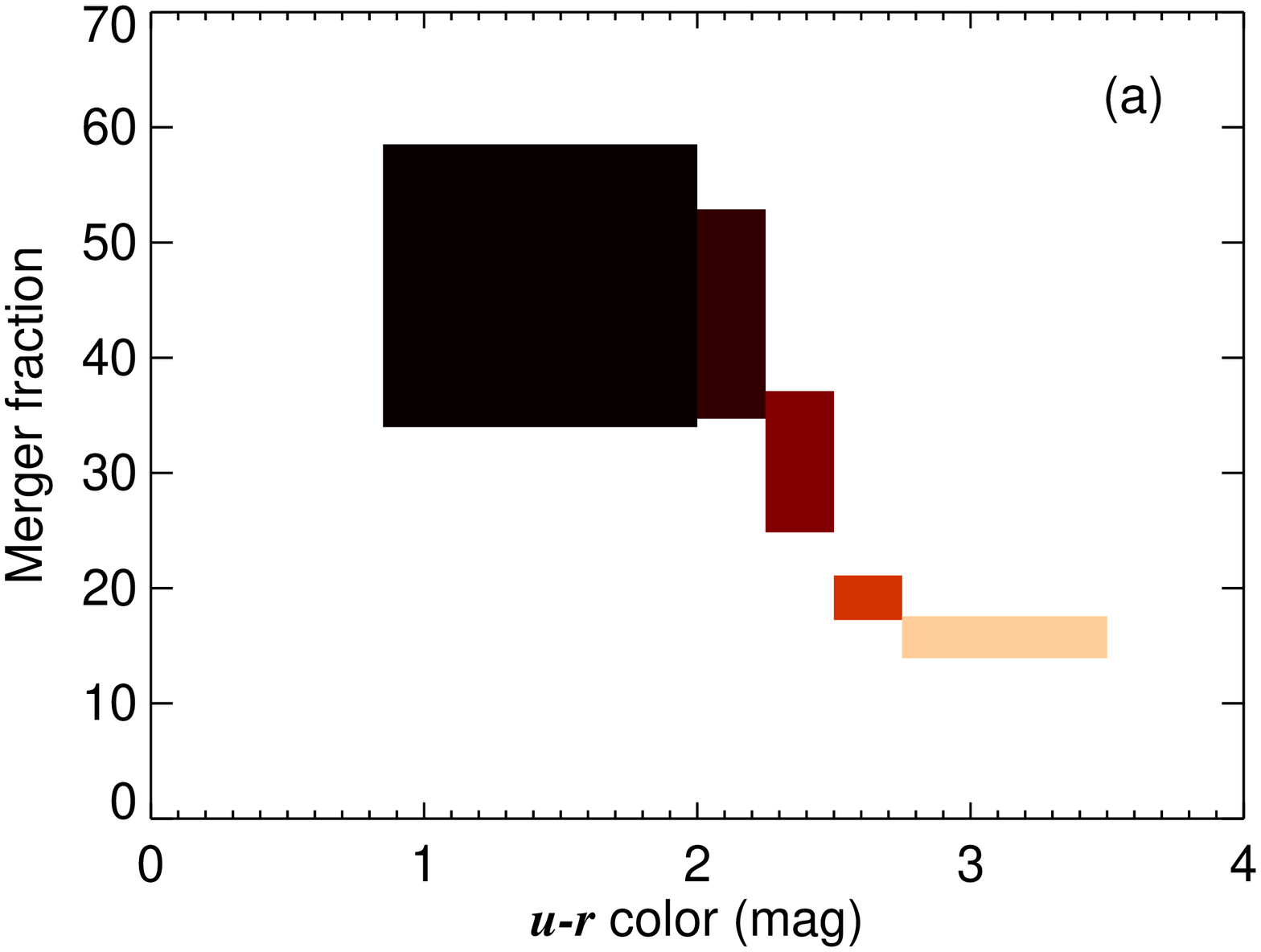}\\
\includegraphics[angle=90, width=0.49\textwidth]{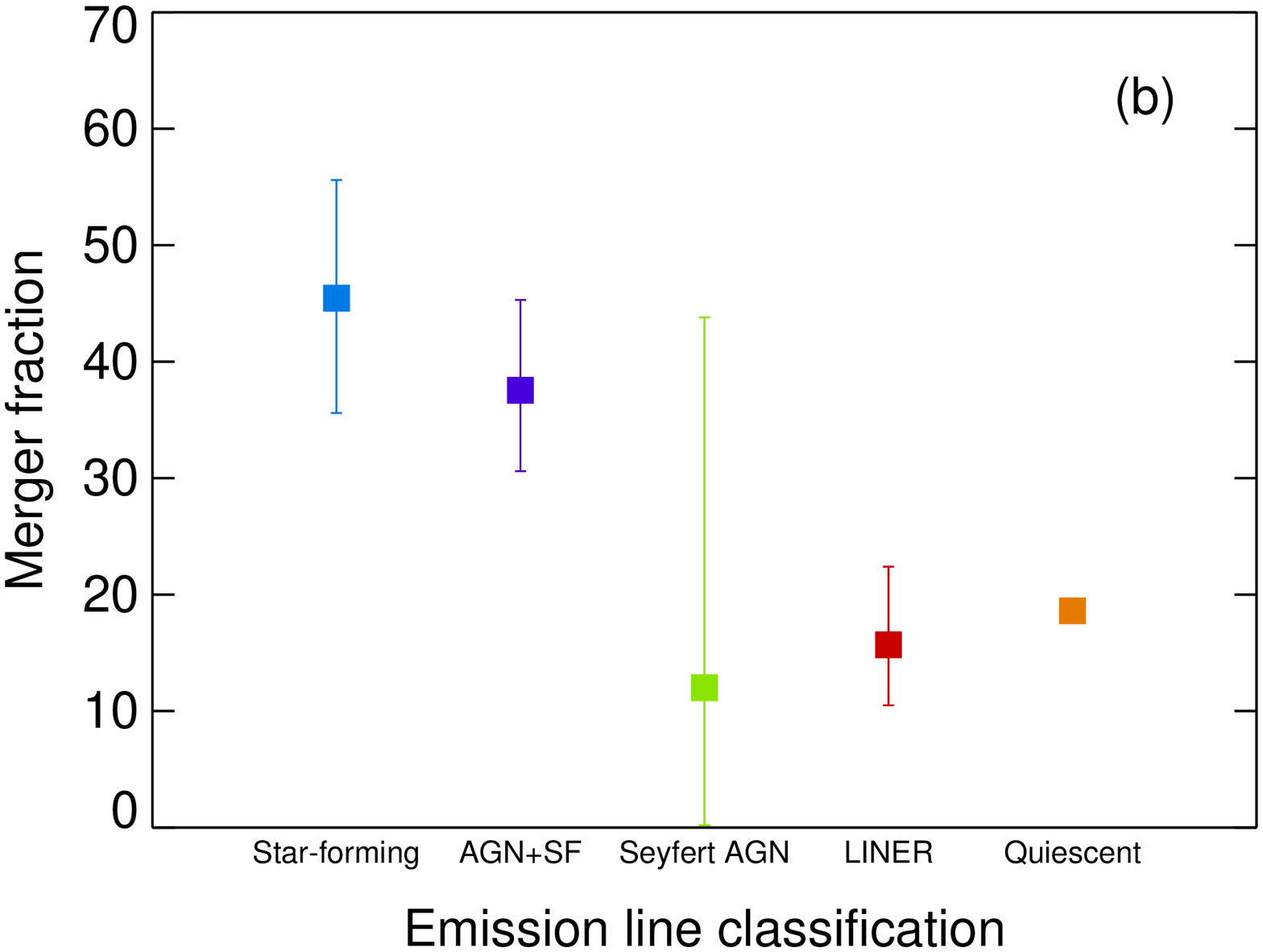}\\
\includegraphics[angle=90, width=0.49\textwidth]{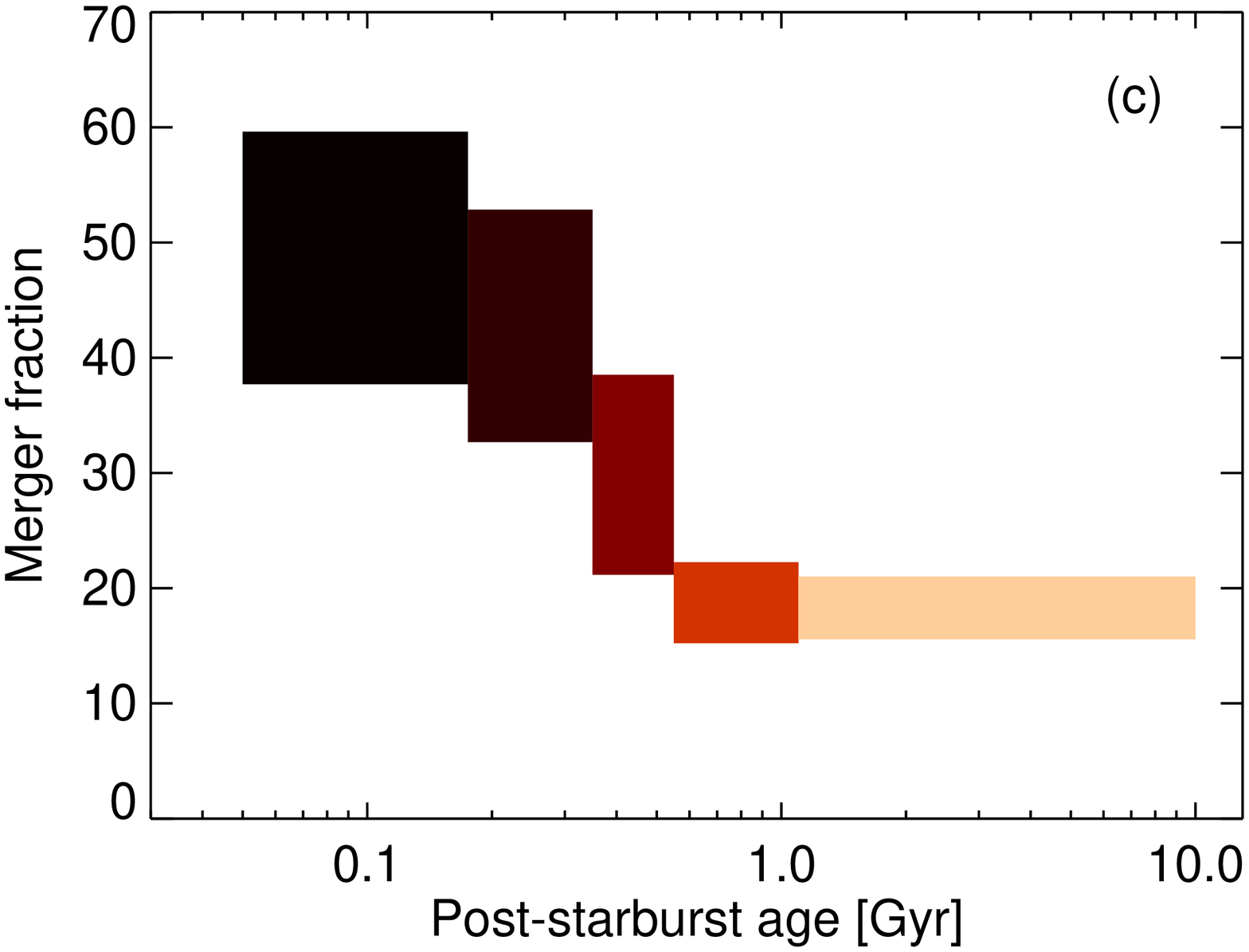}\\
\caption{The merger fraction of low-redshift early-type galaxies transitioning from the blue cloud to the red sequence as a function of various time indicators. \textit{Top}: merger fraction as a function of $u-r$ optical color. \textit{Middle:} merger fraction as a function of emission line classification. \textit{Bottom:} merger fraction as a function of post-starburst age determined from the broad-band SED and stellar absorption indices \citep[see][]{2007MNRAS.382.1415S}. Each emission class is colored differently and labeled below, and in the \textit{top} and \textit{bottom} panels, the bins are selected to sample well the region where the sharp decline occurs. \label{fig:f_merger}}

\end{center}
\end{figure}

\section{Results}

We present our results in Figure \ref{fig:f_merger}. In each panel, we plot the percentage of objects showing evidence for a recent merger as a function of three age indicators. In panel (a), we use the observed optical $u-r$ color as a proxy for age. In panel (b), we instead use the emission line classification and plot the merger fraction as a function of the stages of the evolutionary sequence from star-formation in the blue cloud to quiescence on the red sequence. Finally, in panel (c) we plot the merger fraction as a function of the post-starburst age, $t_{y}$, measured by \cite{2007MNRAS.382.1415S}. The binning of the galaxy sample differs in the three plots; in panel (b), it is by emission line classification, whereas in panels (a) and (c), the color and age bins are chosen to sample the point at which the merger fraction declines.

All three panels in Figure \ref{fig:f_merger} show the same trend: at early times during the migration of early-type galaxies from the blue cloud to the red sequence, the fraction of objects with merger features is high. The merger fraction drops rapidly to the background level of quiescent red sequence galaxies of $\sim20\%$.This drop to the background level of the merger fraction is largely complete by around $500$ Myr after the most recent burst of star formation. This point in time along the evolutionary sequence coincides with the phase during which emission-line-selected Seyfert AGN become prominent.

\section{Discussion}

The decline in merger fraction with age illustrated in Figure \ref{fig:f_merger} has strong implications for the connection between mergers on the one hand and starbursts, black hole accretion and galaxy aging, on the other. We now discuss the implications of the results in Figure \ref{fig:f_merger}:

\subsection{Mergers trigger Starbursts and the Migration from the Blue Cloud to the Red Sequence}
At least half of all blue early-type galaxies show evidence for a recent, significant merger in deeper imaging data. Given that the parent sample of blue early-type galaxies from MOSES \citep{2007MNRAS.382.1415S} was selected for early-type morphology in the shallower regular SDSS images, it is likely that the MOSES sample is missing some objects with young ages ($\sim100$ Myr or less) because they appear disturbed even in the regular SDSS images. This effect would push the merger fraction well above 50\% and may even approach 100\%. This in turn implicates a merger as the trigger for the starburst episode at the beginning of the evolutionary sequence for most, if not all early-types migrating from the blue cloud to the red sequence.

The nature of the merger signature seen in low-redshift early-type galaxies is a question that remains to be investigated. Do the features seen in the deeper images in Figure \ref{fig:sdss_vs_stripe82_merger} necessarily imply a major merger (\textit{i.e.,} a merger with a mass ratio of less than 3:1)? Or can minor mergers produce a similar appearance? Numerical simulations \citep[\textit{e.g.},][]{2008ApJ...684.1062F} suggest that mergers with relatively large mass mass ratios can still produce prominent tidal features. Detailed simulations of mergers and simulated observations mimicking available data are needed to decide whether more information on the nature of the merger can be inferred from our result.

\subsection{Mergers trigger Black Hole Accretion}

Since the Seyfert AGN early-types are consistent with being the post-starburs descendants \citep{2007MNRAS.382.1415S} of the star-forming early-types in the blue cloud, this implies that the merger that triggered the starburst could well have triggered the AGN phase $\sim$500 Myrs later. The time delay between the merger event and black hole feeding could be due to the time required for material to lose sufficient angular momentum to reach the black hole. Alternatively, it may simply be the post-starburst environment, with supernovae and strong stellar winds, that create conditions favorable for accretion onto the black hole.

Our results raise the question of whether the sequence of events that plays out in nature can be accurately described as a ``merger trigger for AGN." It is clear that the time lag between the merger event to the AGN phase is substantial. The merger itself need not be not directly responsible for the delayed AGN phase and in that sense, mergers may not trigger AGN phases. Still, it seems clear that the AGN phase is most likely a consequence of the merger, albeit via a number of intermediate steps. The intermediate steps between the merger and accretion do not remove the causal link between the two events. The question of the merger trigger for AGN phases in early-type galaxies may thus be resolved in an indirect way. 

From a broader perspective, we can also conclude that the evolutionary sequence found by \cite{2007MNRAS.382.1415S} is triggered by mergers. That is, mergers do trigger the migration of low-mass early-type galaxies from the blue cloud to the green valley where black hole feedback processes may be at work advancing this migration \citep{2009MNRAS.396..818S}.

\subsection{A Fundamental Limit to tracing Mergers?}
Another intriguing result of our study is that the rapid decline of the merger fraction to the background level fundamentally limits our ability to link merger signatures over times longer than $\sim500$ Myr. Even though the merger features may survive for much longer, they become indistinguishable from the background level due to numerous dry and minor mergers. Paradoxically, deeper imaging may not extend the detectability or merger feature further in time, as the background merger fraction will also be enhanced in deeper images. There may well be an optimal image depth for studying merger features that maximizes the contrast between background and fading merger features. In order to overcome this limit, a reliable method for separating features due to a recent major mergers and feature induced by the background merger rate would be required.\\

We note that the results presented here are for early-type galaxies \textit{only}. The majority (up to 90\%)  of active nuclei in the local Universe are hosted by late-type galaxies which follow an evolutionary pathway different from the early-types \citep{2010ApJ...711..284S}.

\acknowledgements 
We thank Tomer Tal and Carolin Cardamone for helpful discussions and the anonymous referee for numerous helpful suggestions.

Support for the work of KS was provided by NASA through Einstein Postdoctoral Fellowship grant number PF9-00069 issued by the Chandra X-ray Observatory Center, which is operated by the Smithsonian Astrophysical Observatory for and on behalf of NASA under contract NAS8-03060. KS and CMU gratefully acknowledge previous support from Yale University. Support from NSF grant \#AST0407295 is acknowledged. EE acknowledges support from STFC. 

Funding for the SDSS and SDSS-II has been provided by the Alfred P. Sloan Foundation, the Participating Institutions, the National Science Foundation, the U.S. Department of Energy, the National
Aeronautics and Space Administration, the Japanese Monbukagakusho, the Max Planck Society, and the Higher Education Funding Council for England. The SDSS Web Site is \texttt{http://www.sdss.org/}.

The SDSS is managed by the Astrophysical Research Consortium for the Participating Institutions. The Participating Institutions are the American Museum of Natural History, Astrophysical Institute Potsdam, University of Basel, University of Cambridge, Case Western Reserve University, University of Chicago, Drexel University, Fermilab, the Institute for Advanced Study, the Japan Participation Group, Johns Hopkins University, the Joint Institute for Nuclear Astrophysics, the Kavli Institute for Particle Astrophysics and Cosmology, the Korean Scientist Group, the Chinese Academy of Sciences (LAMOST), Los Alamos National Laboratory, the Max-Planck-Institute for Astronomy (MPIA), the Max-Planck-Institute for Astrophysics (MPA), New Mexico State University, Ohio State University, University of Pittsburgh, University of Portsmouth, Princeton University, the United States Naval Observatory, and the University of Washington.

This research has made use of NASA's Astrophysics Data System Bibliographic Services. \\
{\it Facilities:} \facility{Sloan()}

\bibliographystyle{hapj}


\end{document}